\documentclass{article}
\usepackage{authblk}
\usepackage[utf8]{inputenc}
\usepackage{CJK,multicol,multirow,graphicx,fancyhdr,epstopdf}
\usepackage{amsmath,amsfonts,amssymb,bm,upgreek,mathrsfs,mathcomp}
\usepackage[pagewise,switch,columnwise]{lineno}
\usepackage[compress,nospace]{cite}
\usepackage[dvipsnames]{xcolor}
\usepackage{threeparttable}
\usepackage{booktabs} 
\usepackage[english]{babel}

\usepackage[letterpaper,top=2cm,bottom=2cm,left=3cm,right=3cm,marginparwidth=1.75cm]{geometry}
\usepackage{amsmath}
\usepackage{graphicx}
\usepackage[colorlinks=true, citecolor=blue, allcolors=blue]{hyperref}

\title{Gravitational wave background from extreme-mass-ratio inspirals}

\author[1]{Haoyu Zhao}
\author[4]{Yuanhao Zhang}
\author[1]{Xilong Fan\thanks{xilong.fan@whu.edu.cn}}
\author[2,3,4]{Wenbiao Han}
\affil[1]{School of Physics and Technology, Wuhan University, Wuhan 430072, China}
\affil[2]{Shanghai Astronomical Observatory, Shanghai, 200030, China}
\affil[3]{School of Astronomy and Space Science, University of Chinese Academy of Sciences, Beĳing 100049, China}
\affil[4]{School of Fundamental Physics and Mathematical Sciences, Hangzhou Institute for Advanced Study, UCAS, Hangzhou 310024, China}

\begin{document}
\maketitle

\begin{abstract}
   The gravitational wave background (GWB) produced by extreme-mass-ratio inspirals (EMRIs) serves as a powerful tool for probing the astrophysical and dynamical processes in galactic centers. EMRI systems are a primary target for the space-based detector LISA due to their long-lived signals and high signal-to-noise ratios. This study explores the statistical properties of the GWB from EMRI, focusing on the calculation methods for the GWB, the astrophysical distribution of EMRI sources, and the influence of key parameters, including the spin of supermassive black holes (SMBHs) and the masses of compact objects (COs). By analyzing these factors, we determine the distribution range of the characteristic strain of the GWB from EMRIs. We find that the final eccentricity distributions appear to have negligible effect on the intensity of the GWB due to rapid circularization before they become detectable and the spin of the SMBH enhances the GW characteristic strain by approximately 1$\%$ compared to cases without spin effects. The masses of COs can also significantly affect the characteristic strain of the GWB from EMRIs, with Black Hole (BH) as CO producing a GW signal intensity that is approximately one order of magnitude higher compared to cases where Neutron Star (NS) or White Dwarf (WD) are the COs.
\end{abstract}

\section{Introduction}

In the dense stellar environments of galactic nuclei, there exist EMRIs which consist of a CO and a SMBH. Due to the significant mass disparity between the two objects, typically with the compact object having a mass of $1-50 M_{\odot} $ and the SMBH ranging from $10^4-10^7M_{\odot}$, EMRIs are expected to be strong sources of gravitational waves\cite{AmaroSeoane2010}, especially in the mHz frequency band\cite{Robson2019}, which are the prime targets for the upcoming space-based gravitational wave observatory Laser Interferometer Space Antenna(LISA)\cite{LISA2017}.

Gravitational wave radiation is generated by the changing mass distribution in the binary system\cite{Thorne1980}. In the EMRI system, the main gravitational wave radiation comes from the orbital evolution of small compact body orbiting the SMBH. Since the mass of small compact body is relatively small, its orbit in the gravitational field of the SMBH will follow an elliptical orbit with a large eccentricity. Through dynamical processes such as two-body relaxation\cite{Rauch1996}, resonant relaxation, or interactions with gaseous accretion disks, compact objects are scattered into low angular momentum orbits and become dynamically captured by the central SMBH\cite{Tagawa2020}. Once captured, the EMRI system 
will gradually lose orbital energy via gravitational wave radiation, ultimately leading the compact object to plunge into the SMBH. Due to their slow orbital evolution, EMRIs can remain within the detection band for many years, producing millions of orbital cycles before the final merger\cite{Maggiore2008}. This makes EMRIs excellent candidates for LISA's long-duration observations.

The gravitational wave background is a stochastic signal resulting from the superposition of many unresolved gravitational wave sources across the universe. For EMRI systems, the GWB mainly comes from compact objects orbiting and merging with SMBHs in galactic centers\cite{Renzini2022}, along with contributions from many similar systems in other galaxies. Unlike detecting individual events, the GWB provides an wider view of the population of EMRI sources across the universe, helping us understand the statistical properties and distribution of compact objects in dense galactic centers. This broader perspective allows us to study large-scale astrophysical phenomena, such as the formation and evolution of SMBH, the dynamics of galaxy mergers, and the structure of galactic centers\cite{Christensen2019}.

The study of the GWB from EMRIs represents a critical aspect of GW astrophysics, offering a window into the population and distribution of compact objects in galactic nuclei. Recent research has demonstrated that the statistical properties of the EMRI GWB carry rich information about the density profiles of galactic nuclei, the spin distributions of SMBHs, and the environmental interactions shaping these systems \cite{Barausse2012}. The detection of the EMRI GWB could help resolve long-standing questions about the formation and evolution of SMBHs, as well as the dynamics of the surrounding stellar population\cite{Milosavljevic2003}.
Furthermore, EMRI GWB studies have implications for fundamental physics. The extreme mass ratios and strong gravitational fields near SMBHs provide a natural laboratory for testing general relativity (GR) in the strong-field regime\cite{Barack2007}. Differences from the spacetime geometry predicted by general relativity, such as those caused by alternative gravity theories, may appear as unusual features in the EMRI GWB spectrum. This makes studying the EMRI GWB an essential part of precise gravitational wave astrophysics\cite{Vallisneri2013}.

Key progress in this field includes advancements in modeling the astrophysical distribution of EMRIs and their GW signals. Studies have shown that the spin of supermassive black holes (SMBHs) affects the EMRI GWB, rapidly spinning Kerr black holes enhance GW emission, particularly in higher frequency bands, due to the smaller innermost stable circular orbits (ISCOs) allowed by frame-dragging effects\cite{Hughes2001}. Additionally, eccentricity distributions in EMRI models show that it is important to consider the initial shapes of the orbits accurately, as eccentricity significantly affects the frequency and amplitude of emitted GWs \cite{AmaroSeoane2010}.

Population synthesis studies have provided estimates of the EMRI event rates\cite{Gair2004}, indicating that thousands of EMRI systems could be present within the detection horizon of LISA. This has led to significant efforts to quantify the contribution of unresolved EMRI systems to the stochastic GWB, emphasizing the need for robust detection methods to separate the EMRI background from other GW sources.

The structure of this paper is as follows: in Section 2, we explain how gravitational waves are produced by EMRIs, analyzing the orbital evolution of eccentric binaries and the GW emission from individual sources. Section 3 outlines the methods used to compute the gravitational wave background, including the modeling of EMRI distributions and the calculation of the GWB’s characteristic strain. Section 4 shows and discusses the results, focusing on the characteristic strain of the EMRI GWB under different astrophysical assumptions. Subsections explore the influence of SMBH spins, variations in compact object masses, and the distribution of final orbital eccentricities.  Finally, the conclusion summarizes the main findings and their importance for studying galaxies, black holes, and the early universe.

\section{Gravitational waves from EMRIs}

\subsection{The orbit evolution of eccentric binary}
\begin{figure}
\includegraphics[width=\columnwidth]{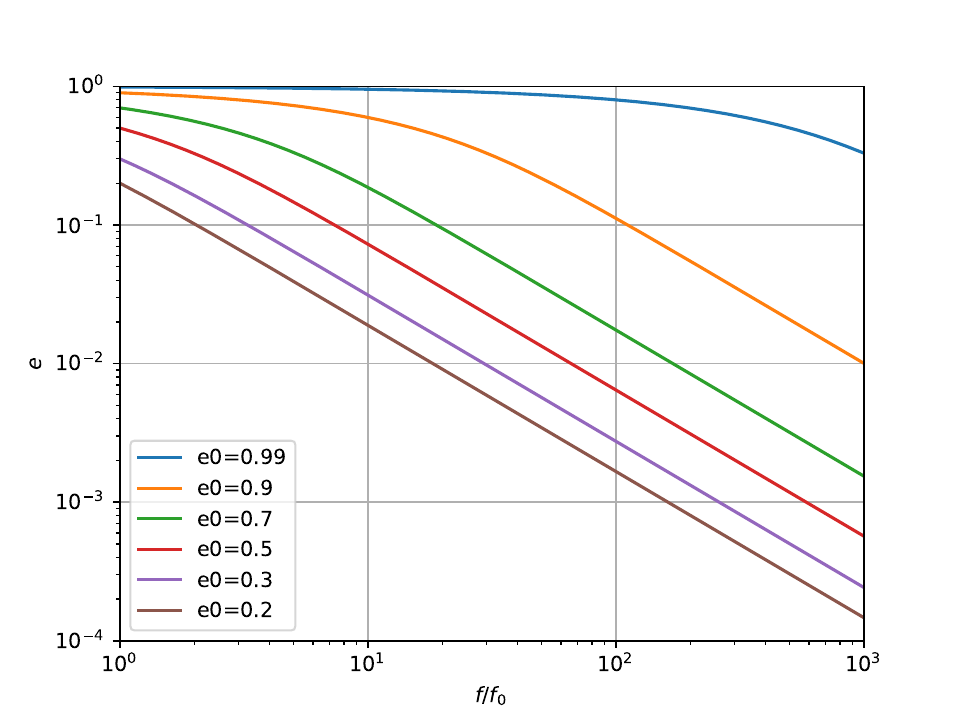}
    \caption{$e-f$  relation. 
    The horizontal axis represents the logarithm of the ratio between the orbital frequency  $f_{\mathrm{orb}}$  and the initial frequency  $f_0$ , while the vertical axis represents the eccentricity  $e$ . Different colors correspond to the evolution in the frequency domain for initial eccentricities  $e_0 = 0.2, 0.3, 0.5, 0.7, 0.9, 0.99$ , respectively. This shows that for a given initial eccentricity  $e_0$ ,  $e$  decreases as the orbital frequency  $f_{\mathrm{orb}}$  increases, and over time, the orbit gradually evolves into a circular one.}
    \label{fig1}
\end{figure}

\begin{figure}
\includegraphics[width=\columnwidth]{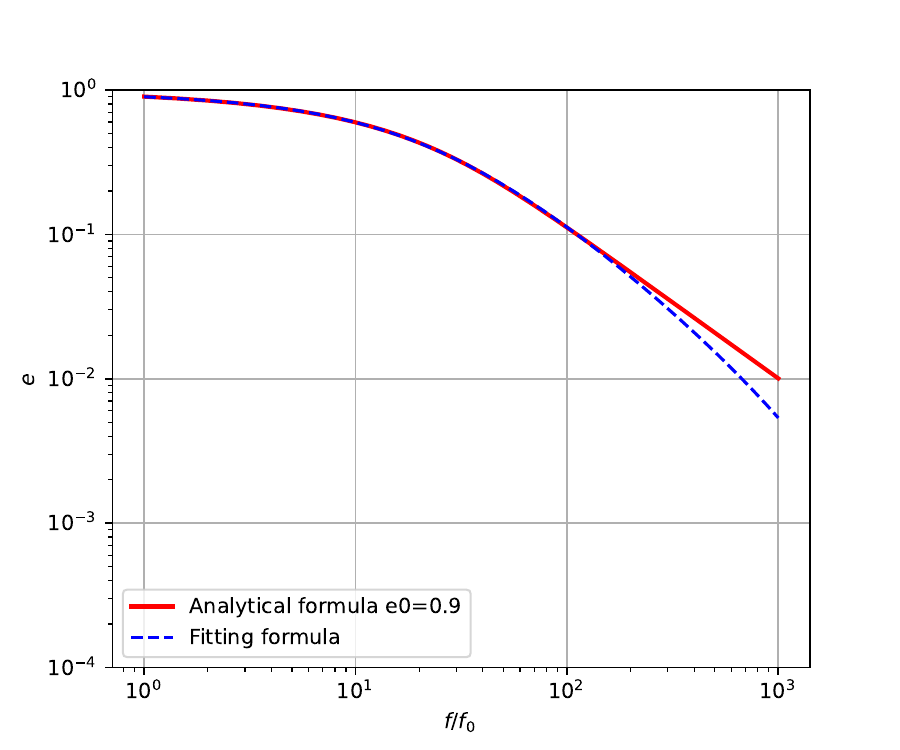}
    \caption{
	Comparison of the fitting function and the analytical function when $e_0=0.9$.The red solid line represents the analytical $e-f$ function, while the blue dashed line represents the interpolated fitting formula. We can see that the fit is relatively good in the low-frequency region, but there is some deviation at higher frequencies.
    }
     \label{fig2}
\end{figure}

Considering a binary system consisting of a massive black hole with mass  $M_1$  and a small compact body with mass  $M_2$, as gravitational waves are radiated, the system continuously loses energy and angular momentum. The evolution of its orbital semi-major axis  $a$  and eccentricity  $e$  is described as follows\cite{Enoki2007}:
\begin{equation}
    \begin{aligned}\label{eq_a-e}
\frac{da}{dt} &=-\frac{64}5\frac{G^3M_1M_2M_{\mathrm{tot}}}{c^5a^3(1-e^2)^{7/2}}\left(1+\frac{73}{24}e^2+\frac{37}{96}e^4\right) \\
\frac{de}{dt} &=-\frac{304}{15}\frac{G^3M_1M_2M_{\mathrm{tot}}}{c^5a^4(1-e^2)^{5/2}}e\left(1+\frac{121}{304}e^2\right)
    \end{aligned}
\end{equation}
where $M_{\text{tot}}=M_1+M_2$. By using Equation (\ref{eq_a-e}), solving the differential equations for ${\mathrm{d}e}/{\mathrm{d}t}$ and ${\mathrm{d}a}/{\mathrm{d}t}$, the relationship between the semi-major axis  $a$  and the eccentricity  $e$  for an elliptical orbit can be obtained:
\begin{equation}
    \frac a{a_0}=\frac{1-e_0^2}{1-e^2}\left(\frac e{e_0}\right)^\frac{12}{19}\left[\frac{1+\frac{121}{304}e^2}{1+\frac{121}{304}e_0^2}\right]^\frac{870}{2299}
\end{equation}
where $a_0$ and  $e_0$  are the initial semi-major axis and initial eccentricity of the orbit, respectively. By further applying Kepler’s Law,  $ a^3\propto f_{\mathrm{orb}}^{-2} $ , the following can be obtained:
\begin{equation}\label{eq3}
    \frac {f_{\mathrm{orb}}}{f_0}=\left\{\frac{1-e_0^2}{1-e^2}\Bigg(\frac e{e_0}\Bigg)^{\frac{12}{19}}\Bigg[\frac{1+\frac{121}{304}e^2}{1+\frac{121}{304}e_0^2}\Bigg]^{\frac{870}{2299}}\Bigg\}^{-3/2}\right.
\end{equation}
where  $f_0$  is the initial orbital frequency. It should be noted that the evolution of the eccentricity  $e$  as a function of the orbital frequency  $f_{\mathrm{orb}}$ ,  $f(e)$ , cannot be obtained analytically. Therefore, polynomial interpolation is required for the calculation. Figure \ref{fig1} shows the  $f(e)$  curves obtained through interpolation for different initial eccentricities.
It can be seen that as the frequency increases, the orbital eccentricity gradually approaches zero, and the elliptical orbit evolves into a circular orbit. In the subsequent calculations, we use the interpolation approximation from Yunes' article\cite{Yunes2009}.：
\begin{equation}\label{eq4}
    e(f)=\frac{16.83-3.814\mathrm{~}\beta^{0.3858}}{16.04+8.1\mathrm{~}\beta^{1.637}}
\end{equation}
where  $\beta = \chi^{2/3} / \sigma_0$ ,  $\chi = f_{\mathrm{orb}} / f_0$ ,  $f_{\mathrm{orb}}$  is the orbital frequency.
\begin{equation}
    \sigma(e_0)=\frac{e_0^{12/19}}{(1-e_0^2)}\left[1+\frac{121}{304}e_0^2\right]^{870/2299}
\end{equation}
 As shown in Figure\ref{fig2}, we compare the analytical $e\text{-}f$ relation (red solid line) with the interpolated fitting formula (blue dashed line) for an initial eccentricity $e_{0} = 0.9$. At lower orbital frequencies \(\bigl(f_{\mathrm{orb}}/f_{0} \lesssim 10^2\bigr)\), the two curves coincide quite closely. This agreement indicates that, in the early stages of inspiral, the interpolation captures the key behavior of eccentricity damping accurately.

\subsection{The GW emission from a single EMRI source}
\begin{figure}
\includegraphics[width=\columnwidth]{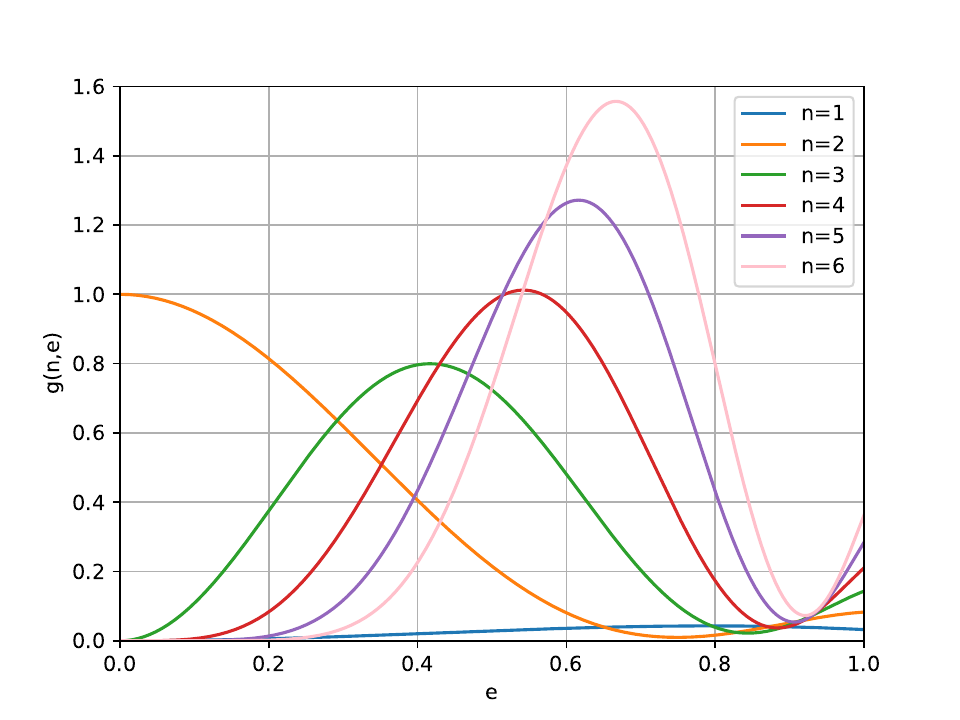}
    \caption{The function $g(n,e)$ evolves with eccentricity $e$ for harmonic numbers \( n = 1, 2, 3, 4, 5, 6  \), where \( g(2, 0) = 1 \) when \( n = 2 \) and \( e = 0 \), representing a circular orbit. The graph illustrates that higher-order harmonics contribute more significantly to the total gravitational wave radiation as eccentricity increases.}
    \label{fig3}
\end{figure}

For an EMRI system composed of masses $M_1$ and $M_2$ , the total gravitational wave radiation power distributed across different harmonics is represented by $\dot{E}_n$ \cite{Huerta2015}：
\begin{equation}\label{eq6}
\dot{E}_n(M_1,M_2,f_{\mathrm{orb}},e)=\dot{E}_{\text{circ}}g(n,e)
\end{equation}
where $\dot{E}_{\text{circ}}$ represents the gravitational wave radiation power for a circular orbit\cite{Peters1963}:
\begin{equation}
\dot{E}_{\text{circ}}(M_1,M_2,f_{\mathrm{orb}})=\frac{32G^{7/3}}{5c^5}(2\pi\mathcal{M}f_\mathrm{orb})^{10/3}
\end{equation}
in the equation,  $\mathcal{M}$ refers to the chirp mass:
\begin{equation}
 \mathcal{M}=\frac{(M_1M_2)^{3/5}}{(M_1+M_2)^{1/5}}
\end{equation}
The latter part of Equation (\ref{eq6}),  $g(n,e)$  is a dimensionless distribution function that reflects the contribution of the orbital eccentricity  $e$  and the harmonic number  $n$  to the gravitational wave radiation power from an elliptical orbit.
\begin{equation}
    \begin{aligned}
g(n,e)& =\frac{n^4}{32}\bigg[ \bigg\{J_{n-2}(ne)-2eJ_{n-1}(ne) \\
&+\frac{2}{n}J_n(ne)+2eJ_{n+1}(ne)-J_{n+2}(ne)\bigg\}^2 \\
&+\left(1-e^2\right)\bigg\{J_{n-2}(ne)-2J_n(ne)+J_{n+2}(ne)\bigg\}^2 \\
&+\frac{4}{3n^2}J_n^2(ne)\bigg].
\end{aligned}
\end{equation}
Here,  $J_n$  represents the Bessel function of the first kind of order  $n$. Figure (\ref{fig3}) shows the evolution of  $g(n, e)$  as a function of eccentricity  $e$  for different harmonic orders. We can observe that as eccentricity increases, the gravitational wave radiation contribution from higher-order harmonics becomes more significant in the total gravitational wave radiation.
Now, consider the total power radiated by this source in gravitational waves $\dot E$: 
\begin{equation}\label{eq10}
 \dot E= \frac{32G^{7/3}}{5c^5}(2\pi\mathcal{M}f_\mathrm{orb})^{10/3}\sum_{n=1}^{\infty}g(n,e)
\end{equation}
 then transform the power spectrum of the gravitational waves into the frequency domain:
\begin{equation}
    \frac{dE}{df}=\frac{\pi^{2/3}G^{2/3}\mathcal{M}^{5/3}}{3F(e)}f^{-1/3}\label{E(f)}
\end{equation}
the results from Peter\cite{Peters1963} can be utilized to simplify the subsequent summation over $g(n,e)$:
\begin{equation}
F(e)=\sum_{n=1}^\infty g(n,e)=\frac{1+\frac{73}{24}e^2+\frac{37}{96}e^4}{\left(1-e^2\right)^{7/2}}
\end{equation}
Finally, we introduce the characteristic amplitude  $h_{c,n}$  to describe the strength of the gravitational waves. Since it is directly related to the energy density of the gravitational wave, it represents the average amplitude of the wave at a specific frequency. Its definition is given as follows\cite{Finn2000}:
\begin{equation}\label{eq12}
    h_{c,n}=\frac1{\pi d}\sqrt{\frac{2G\dot{E}_n}{c^3\dot{f}_n}}
\end{equation}
 here $f_n=nf_{\mathrm{orb}}$  denotes the GW frequency of the $n$-th harmonic in the source’s rest frame, where  $f_{\mathrm{orb}}$  is the orbital frequency. $\dot E_n$  represents the gravitational wave radiation power at the $n$-th harmonic, and  $d$  is the comoving distance. The time derivative of  $f_\text{orb}$  can be derived from Equation (\ref{eq_a-e}) and Kepler's third law $ a^3\propto f_{\mathrm{orb}}^{-2} $ , as follows: 
\begin{equation}\label{eq13}
    \frac{df_\mathrm{orb}}{dt}=\frac{96G^{5/3}}{5c^5}(2\pi)^{8/3}\mathcal{M}^{5/3}f_\mathrm{orb}^{11/3}{F}(e)
\end{equation}
By substituting Equations (\ref{eq10}) and (\ref{eq13}) into Equation (\ref{eq12}), the expression for the characteristic amplitude $h_c$ can be obtained：
\begin{equation}
    h_c^2=\sum_{n=1}^\infty h_{c,n}^2=\frac{2G^{5/3}\pi^{2/3}\mathcal{M}^{5/3}f^{-1/3}}{3c^3\pi^2d^2}\times \Phi(f)
\end{equation}
where $\Phi(f)$ is the function representing the influence of eccentricity on the GW radiation power, and it is defined as:
\begin{equation}
    \Phi(f)=\sum_{n=1}^\infty\Phi_n=\sum_{n=1}^\infty \left(\frac2n\right)^{2/3}\frac{g(n,e)}{F(e)}
\end{equation}
In Huerta's paper\cite{Huerta2015}, a detailed analysis of the properties of  $\Phi$  is provided, with the following key characteristics:
\begin{itemize}
    \item The frequency ratio $f^{\max}/f_0$  corresponding to the peak value of $\Phi(f)$  is given by:
    \begin{equation}
        \frac{f^{\max}}{f_{0}}=\frac{1293}{181}\left[\frac{e_0^{12/19}}{1-e_0^2}\left(1+\frac{121e_0^2}{304}\right)^{870/2299}\right]^{3/2}
    \end{equation}
    \item For all $e_0$ , the maximum value of  $\Phi$  is $\Phi_{\max}={373}/{234}$;
    \item When the initial eccentricity  $e_0$  is 0, $\Phi=1$;
    \item As the frequency  $f$  approaches infinity,  $\Phi$  tends toward 1.
\end{itemize}

\begin{figure}
    \includegraphics[width=\columnwidth]{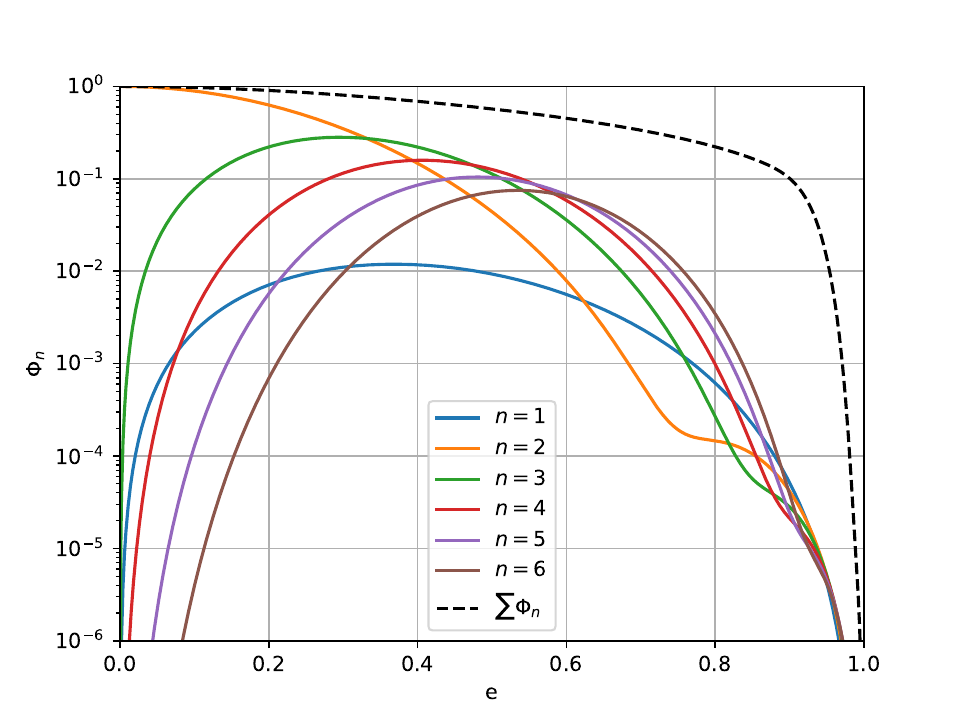}
    \caption{ The solid lines represent the distribution of $ \Phi_n $ as a function of eccentricity $ e $ for harmonic orders $ n = 1, 2, 3, 4, 5, 6 $. As $ \Phi $ lacks an analytical solution, it must be determined through numerical methods or polynomial interpolation. In the range $ 0 < e < 0.9 $, the interpolation provides a good fit, allowing the use of a fitting function to approximate the values and thus significantly reduce computational complexity.}
    \label{fig4}
\end{figure}

\section{GW Background Calculation}
\subsection{The distribution of EMRIs }

\begin{figure}
        \centering        \includegraphics[width=1\linewidth]{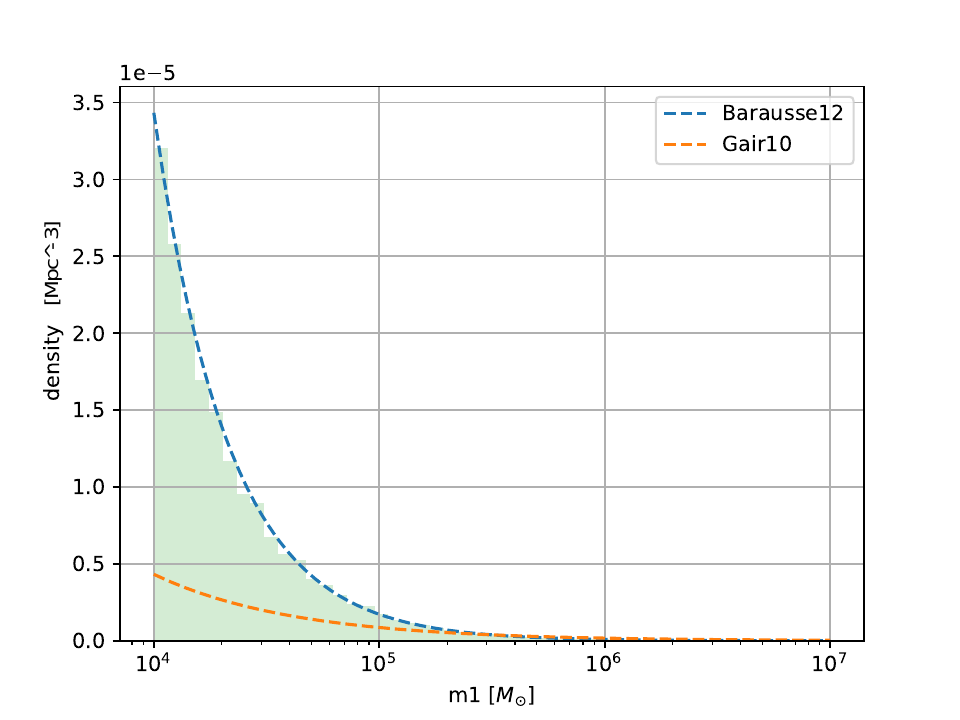}
        \caption{This figure shows two different SMBH mass distribution models. The blue dashed line represents the B12 model, which corresponds to an optimistic scenario, while the orange dashed line corresponds to the G10 model, representing a lower threshold. The histogram represents the mass distribution we fitted in model 1. Plotting histograms can help us verify the accuracy of the fitted data by providing a visual comparison between the model predictions and the data distribution.}
        \label{fig:5}
    \end{figure}

We utilize the EMRI's population models provided by Babak\cite{Babak2017} to  simulate the EMRI rates $\text{d}N/\text{d}t$. The specific parameters are listed in Table 1. We selected these three representative models: M1 as a fiducial model, M11 as a pessimistic model, and M12 as an optimistic model. Each EMRI population model assumes an observation time of 10 years and considers distances up to a redshift of  $z = 4.5$ .

The Table 1 shows two  popular models for describing the mass distribution of SMBHs , which are the Gair10 and Barausse12 models, each with unique assumptions and characteristics that affect the resulting EMRI predictions.

The Gair10 model assumes that the mass distribution of SMBHs is primarily influenced by galaxy mergers, with a specific focus on SMBHs associated with galactic nuclei. It predicts a lower number of low-mass SMBHs, which aligns with the observed galaxy evolution. It provides an essential basis for simulating EMRI rates under the assumption that SMBH mass does not vary dramatically over cosmic time. This model assumes that SMBHs follow a simple mass distribution, represented by the formula:
\begin{equation}
    \frac{dn}{d\log M}=0.002\left(\frac{M}{3\times10^{6}M_{\odot}}\right)^{0.3}\mathrm{~Mpc}^{-3}
\end{equation}
The Barausse12 model aligns more closely with the current astrophysical understanding of MBH formation and evolution, and thus finds broader application in predicting EMRI events. Its dynamic evolutionary features allow it to provide more accurate forecasts of EMRI event rates, especially across different redshift and mass ranges.
it can be expressed as following formula:
\begin{equation}
    \frac{\mathrm{d}n}{\mathrm{d}\log M}=0.005\left(\frac{M}{3\times10^{6}M_{\odot}}\right)^{-0.3}\mathrm{~Mpc}^{-3}
\end{equation}
These two models are shown in the Figure \ref{fig:5}, we can find that the B12 model is an optimistic scenario and the G10
model represents a lower threshold which is a pessimistic scenario.

To calculate the GWB characteristic strain which is derived from the superposition of many unresolved gravitational wave sources, it's essential to consider the contributions from EMRI systems at each frequency\cite{Barack2009}. Therefore, we need to know the number of EMRI sources near a specific frequency $f$ to estimate the gravitational wave strength contributed by all sources in that frequency range $dN/df$ by using following relationship:
\begin{equation}
    \frac{dN}{df} = \frac{dN}{dt} \frac{dt}{df}
\end{equation}
where $dt/df$ can be obtained in equation (\ref{eq13}). For EMRIs, their distribution involves four relevant properties: mass, redshift, initial orbital eccentricity, and orbital frequency, which can be expressed as $d^4N/d\mathcal{M}\  dz \ de \ df_{\mathrm{orb}}$. To gain the list, we randomly sample points over the observation period and integrate backward from the final eccentricity and semi-major axis, recording orbital frequency and eccentricity at each point to represent the different stages in the EMRI evolution.
\begin{table*}
    \centering
    \begin{tabular}{lccccccc|c}
        \toprule
        Model & Mass function & MBH spin & Cusp erosion & $M$-$\sigma$ relation & $N_p$ & CO mass $[M_\odot]$ & \multicolumn{1}{c|}{EMRI rate [yr$^{-1}$]} \\
        \midrule
        M1 & Barausse12 & a98 & yes & Gultekin09 & 10 & 10 & 1600  \\
        M11 & Gair10 & a0 & no & Gultekin09 & 0 & 10 & 13  \\
        M12 & Barausse12 & a98 & yes & Gultekin09 & 0 & 10 & 20000 \\
        \bottomrule
    \end{tabular}
    \caption{EMRI models provided by Babak et al. \protect\cite{Babak2017} with various parameters and corresponding rates of EMRI detections. The first column defines the models' label, column 2 shows the mass function of each model, while column 3 presents the MBH spin model. Column 4 indicates whether the effect of cusp erosion is included. Column 5 lists the  $M\text{–}\sigma$  relation used in each model. Column 6 provides the ratio of plunges to EMRIs. Column 7 specifies the mass of the compact objects (COs), and column 8 gives the total EMRI merger rate (in yr$^{-1}$) up to redshift  $z = 4.5$ .}
    \label{tab:emri_models}
\end{table*}

\subsection{Calculation of the GW Background}
In any homogeneous and isotropic universe, the energy density  $\rho_{\text{GW}}$  of the gravitational wave background radiation, considering redshift effects, should sum the radiation energy density from all sources at each redshift  $z$ , as follows:
\begin{equation}\label{eq17}
\rho_\text{GW}c^2=\int_0^\infty\int_0^\infty dz d\mathcal{M} \frac{d^2 n_c}{dz d\mathcal{M} }\frac1{1+z}\frac{dE_\text{GW}}{ d f _ r }df_r
\end{equation}
Here,  $n_c$  represents the comoving number density of gravitational wave sources within a given comoving volume at redshift range  $z \sim z+dz$ .  $f_r$  is the frequency of the gravitational wave in the rest frame of the source, and  $f$  is the frequency in the observer’s frame. According to the redshift effect, they satisfy  $f_r = f(1+z)$ .
The quantity $dE_\mathrm{GW}$ can be written as:
\begin{equation}\
    dE_\mathrm{GW}=\frac{dE_\mathrm{GW}}{df_r}df_r
\end{equation}
which represents the total energy radiated by the source in gravitational waves within the frequency range  $f_r \sim f_r + df_r$ . This total energy is measured in the source’s rest frame and is integrated over the entire lifetime of the source and all solid angles. According to the expression provided in \cite{Phinney2001}, the total gravitational wave energy density is given by:
\begin{equation}\label{eq18}
    \rho_\text{GW}c^2=\int_0^\infty\frac{\pi}{4}\frac{c^2}{G}f^2h_c^2(f)\frac{df}{f}
\end{equation}
From Equations (\ref{eq17}) and (\ref{eq18}), we can derive the formula for calculating the characteristic strain of the gravitational wave background:
\begin{equation}
h_{c,\text{gwb}}^2( f) =\frac{4G}{\pi c^2f^2}\int_0^\infty dz \, d\mathcal{M} \frac{d^2 n_c}{dz \, d\mathcal{M} }\frac1{1+z}.\left(f_r\frac{dE_\mathrm{GW}}{df_r}\right)
\end{equation}
We can express the comoving number density of EMRIs as the comoving number of binaries emitting in a given logarithmic frequency interval with chirp mass and redshift
in the range ($\mathcal{M},\mathcal{M}+d\mathcal{M}$) and ($z,z+dz$):
\begin{equation}
    \frac{d^2n_c}{dz d\mathcal{M}}=\frac{d^3N}{dz \, d\mathcal{M} \, d\mathrm{ln}f_\mathrm{orb}}\frac{d\mathrm{ln}f_\mathrm{orb}}{dt_r}\frac{dt_r}{dz}\frac{dz}{dV_\mathrm{c}}
\end{equation}
$dV_c$ is the comoving volume shell lying between $z$ and $z + dz$, we can gain:
\begin{equation}
\frac{d^2n_c}{dz \, d\mathcal{M}} = \frac{1}{(1+z) 4\pi c d^2} \frac{d^3N}{dz \, d\mathcal{M} \, d\ln f_{\mathrm{orb}}} \frac{d\ln f_{\mathrm{orb}}}{dt_r}
\end{equation}

Although the eccentricity $e$ does not explicitly appear in the basic formula for calculating the GW background, it affects the residence time and radiation intensity at each frequency, and influencing the overall distribution of the GW background. To accurately estimate the GW background , it is essential to consider the initial eccentricity distribution of the sources. By  substituting Equation (\ref{E(f)}) and (\ref{eq12}) into the expression for $h_{c,\mathrm{gwb}}$ and taking into account the eccentricity distribution of EMRI sources, we obtain:
\begin{equation}    
h_{c,\mathrm{gwb}}^2(f)=\int dz \int d\mathcal{M} \int de  \left[\sum_n\frac{\mathrm{d}^4N}{dz d\mathcal{M} de \mathrm{d}\ln f_{\mathrm{orb}}}h_n^2(f)\right]_{f_{\mathrm{orb}}=\frac{f(1+z)}{n}}
\end{equation}
Here,  $h_n$  is the root mean square (rms) strain of the  $n$-th harmonic \cite{Finn2000}. Through the following conversion relationship, $h_n$ can be replaced by $h_{c,n}$:
\begin{equation}
    h_{c,n}= h_{n}\sqrt{\frac{2f_n^2}{\dot{f}_n}}
\end{equation}
and we sum over all harmonics:
\begin{equation}\label{eqhc}
    h_{c,\mathrm{gwb}}^2(f)=\frac{1}{2}\int dz \int d\mathcal{M} \int de  \bigg[\sum_{n=n_{\mathrm{min}}}^{n_{\mathrm{max}}}\frac{d^4N}{dz d\mathcal{M} de d\ln f_{\mathrm{orb}}}\frac{h_{c,n}^2(f)}{fT_{\mathrm{obs}}}\bigg]_{f_{\mathrm{orb}}=\frac{f(1+z)}{n}}
\end{equation}
where the range of orbital frequency depends on the orbital frequency at beginning of the observation $f_{\text{orb}}(t=0)$ and the orbital frequency at the end of the observation $f_{\text{orb}}(t=T_{\text{obs}})$ or at binary coalescence $f_{\text{orb}}(t=t_{\text{ISCO}})$.

To calculate this integral, We begin by assigning each EMRI system a final eccentricity $e_{\mathrm{final}}$ between 0 and 0.2 (either uniform or following a power-law). Next, we use the orbital evolution equations (e.g., Peters’ formula) to integrate backward in time from $e_{\mathrm{final}}$, determining the initial eccentricity $e_0$. We similarly estimate the final semi-major axis $a_{\mathrm{final}}$ near the ISCO and integrate backward to find the corresponding initial semi-major axis $a_0$. Repeating this procedure for all EMRIs produces an array of $(e, a)$ values for each system. We then set a finite time step and integrate forward in time over a chosen observation window (e.g., 4 years), tracking how $a(t)$ and $e(t)$ change due to gravitational-wave emission. At each time step, we compute the orbital frequency $f_{\mathrm{orb}}(t)$, the associated GW frequency $f$, and the characteristic strain $h_c$. Finally, we sum the strain contributions $h_c$ from all EMRIs at each frequency bin to obtain the total GWB spectrum $h_c(f)$. This procedure ensures that orbital circularization, from the initial eccentricity $e_0$ to values near zero, is accounted for within the inspiral window relevant to the detector.

\section{Results and Discusion}
In this work, we have systematically investigated the characteristic strain of the gravitational wave background arising from EMRIs under various assumptions about the spin of the SMBH, the final eccentricity distribution, the mass distribution models of EMRIs, and the mass range of compact object companions. Our analysis includes a comparison across multiple models to assess the relative influence of each factor on the GWB signal, focusing on frequency-dependent strain characteristics.

\begin{figure}
 \includegraphics[width=\columnwidth]{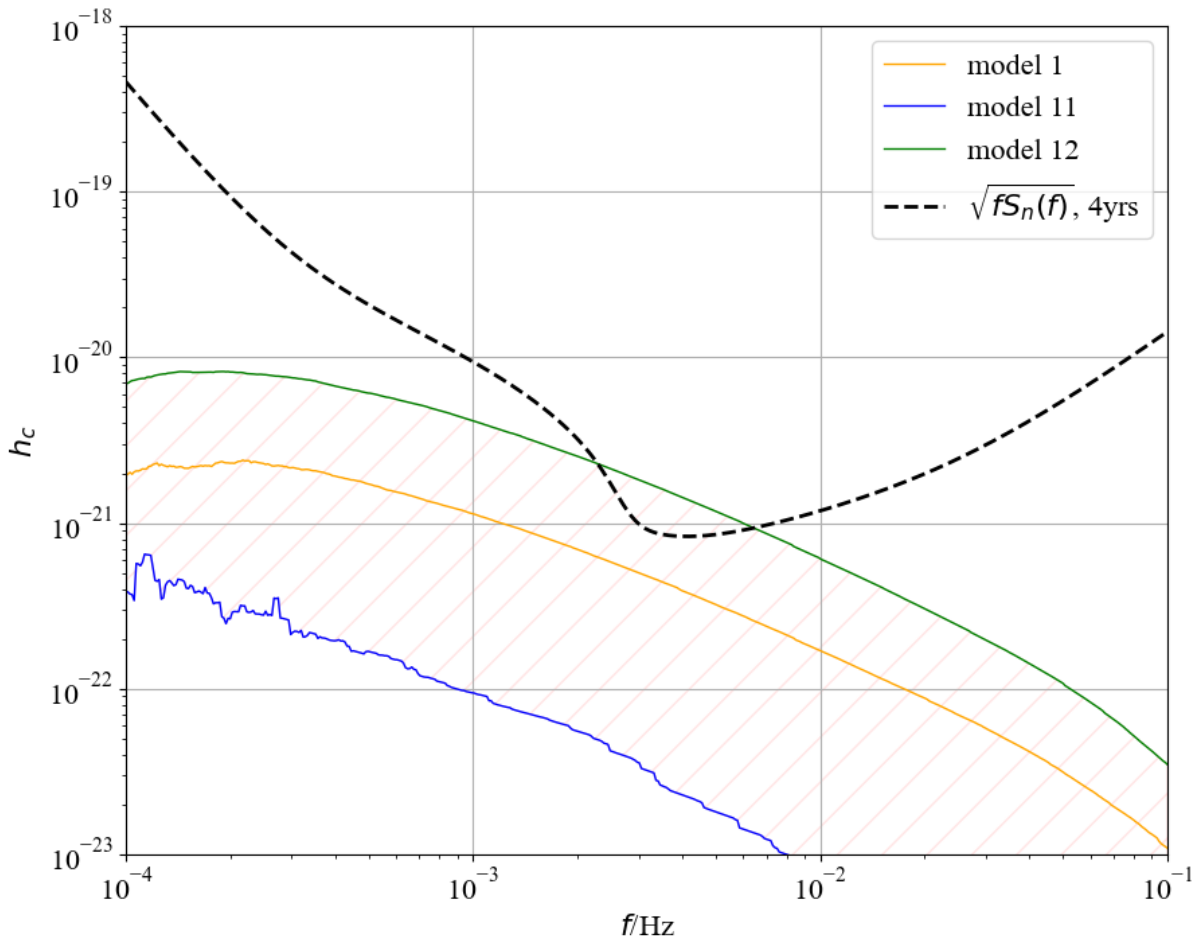}
    \caption{ This figure shows the variation of the GWB characteristic strain with frequency for three different EMRI models. Model 1 is the fiducial model, model 11 represents the pessimistic scenario, and Model 12 is the optimistic scenario. The pink shaded region represents the distribution range of $h_c$. The black dashed line represents the LISA's sensitivity curve ($ T_{\text{obs}} = 4 \, \text{yr} $).}
    \label{fig 5}.
\end{figure}

The GWB spectrum can be constructed as the sum of individual source strains from all binaries emitting at the relevant frequencies and harmonics within the observation time. By using 
Monte Carlo method, we can obtain the results in figure(\ref{fig 5}), in which there are three curved lines representing the optimistic, pessimistic, and fiducial models. The black dashed line
represents the LISA’s sensitivity curve when the observation time is 4 years. We find that the optimistic model lies above LISA’s sensitivity curve which can be detected in the frequency range $10^{-3} - 10^{-2}$Hz.

\begin{figure}
\includegraphics[width=\textwidth]{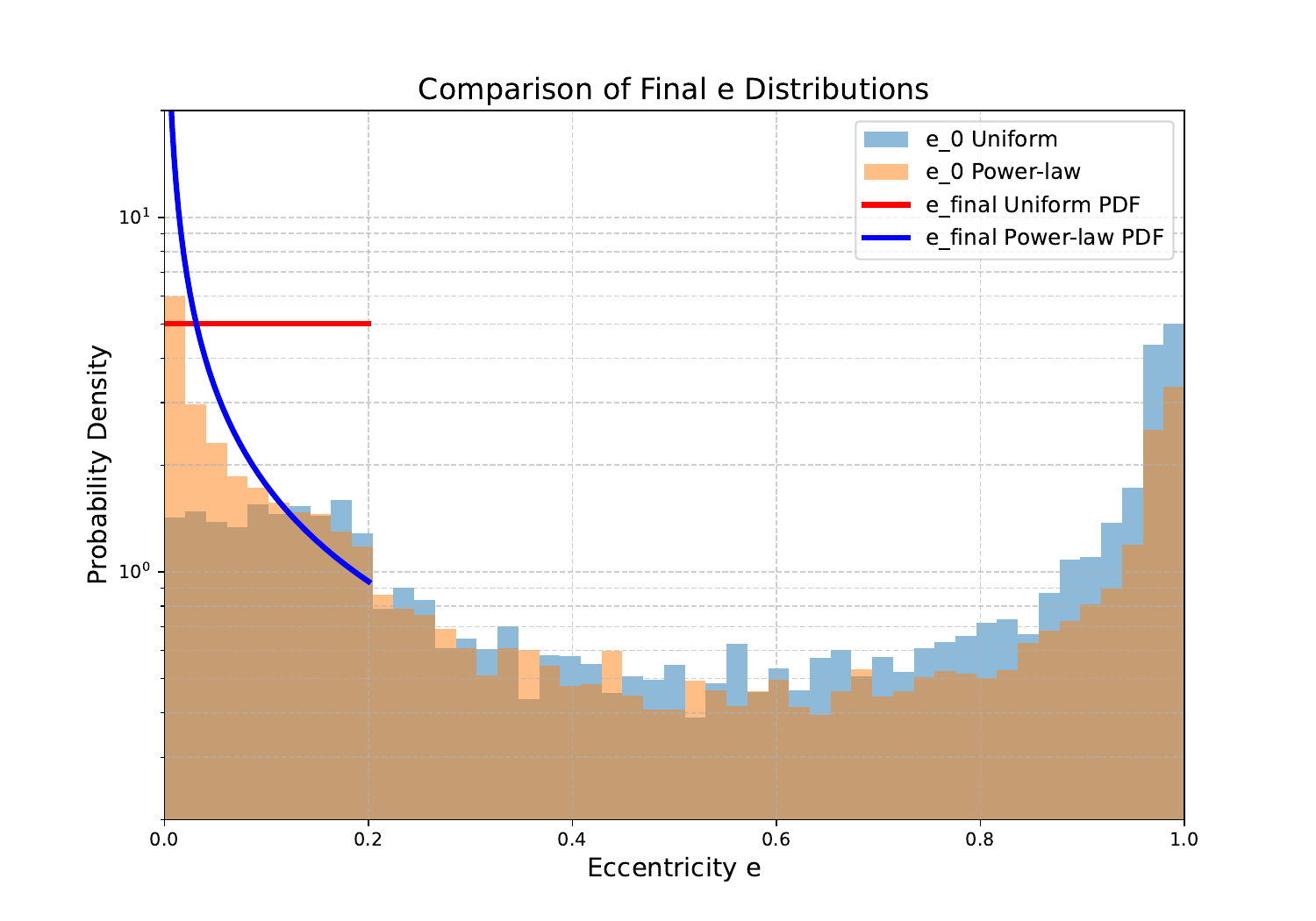}
\caption{ This figure illustrates the comparison between the final eccentricity distributions and the corresponding initial eccentricity $e_0$ distributions for EMRIs, with final eccentricities restricted to the range 0–0.2. The solid curves represent the final eccentricity distributions after the systems have evolved under GW emission. Overlaid on these curves are histograms representing the initial $e_0$ distributions: the brown histogram is derived from a power-law $e_{\text{final}}$ distribution, and the blue histogram comes from a uniform $e_{\text{final}}$ distribution.
}\label{e0_ef}
\end{figure}

\begin{figure}
\includegraphics[width=1\linewidth]{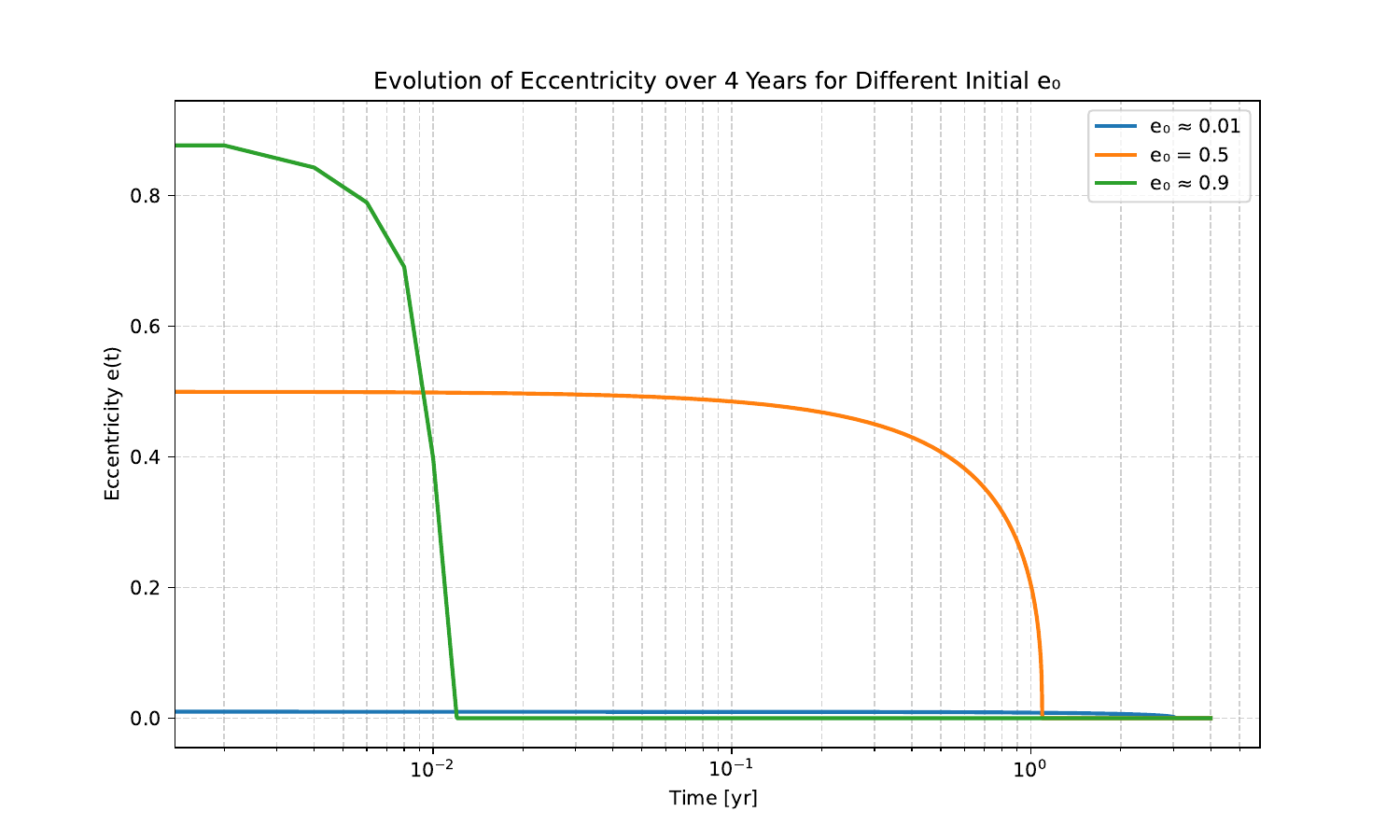}
  \caption{Evolution of orbital eccentricity e(t) over 4 years for EMRI systems with different initial eccentricities. The blue, orange, and green curves correspond to initial eccentricities $e_0 \approx 0.01$, $e_0 = 0.5$, and $e_0 \approx 0.9$, respectively. Systems with very high $e_0$ exhibit rapid circularization within a short timescale ($\sim$ 0.01 yr), while those with intermediate or low initial eccentricities evolve more gradually. This graph illustrates the strong eccentricity damping effect of gravitational wave radiation, especially at high eccentricities.}
  
  \label{e-t}
\end{figure}

We calculate and plot the EMRI GWB for two different final eccentricity distributions.As shown in figure(\ref{e0_ef}).The distribution of initial eccentricities $e_0$ for EMRI systems typically exhibits a bimodal shape. A high-eccentricity peak (around 0.8–1) arises naturally from standard EMRI formation mechanisms such as two-body relaxation and dynamical scattering, where stellar-mass objects are captured by the MBH on highly eccentric orbits.
On the other hand, a low-eccentricity peak (around 0–0.2) can be attributed to the strong circularizing effect of gravitational wave emission. Due to the nonlinear nature of energy and angular momentum loss, systems with very high eccentricities can undergo rapid circularization once they enter the LISA band. For example, as shown in Fig \ref{e-t}, a system with $e_0 = 0.9$ can evolve to a nearly circular orbit (e.g., $e \sim 0.01$) within a very short timescale ($\sim0.01$yr). This rapid evolution can map some systems that were originally formed with high eccentricities into the low-eccentricity region by the time they enter the observable window, giving rise to the second peak at low $e_0$ when inferred from final observed values via backward evolution.

\begin{figure}
\includegraphics[width=1\linewidth]{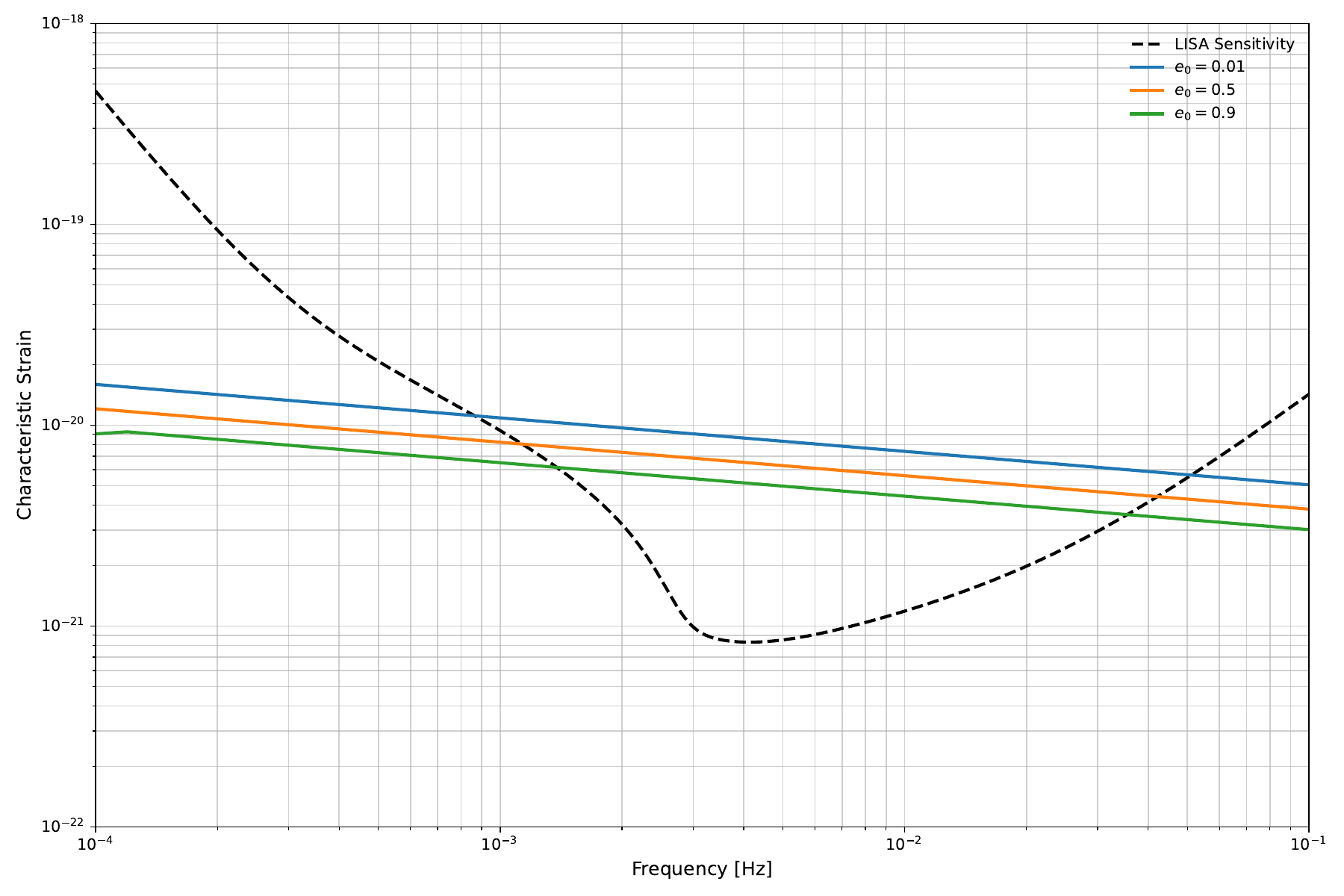}
\caption{ Characteristic strain spectra $h_c(f)$ for EMRI systems with different initial eccentricities $e_0$ = 0.01, 0.5, and 0.9 for single sources at $z=4.5$, shown against the LISA sensitivity curve (dashed line). Despite the variation in initial eccentricity, the resulting strain curves lie close to each other across the LISA-sensitive frequency band, with differences of at most a few factors. }
\label{fig9}
\end{figure}

\begin{figure}
    \centering    \includegraphics[width=1\linewidth]{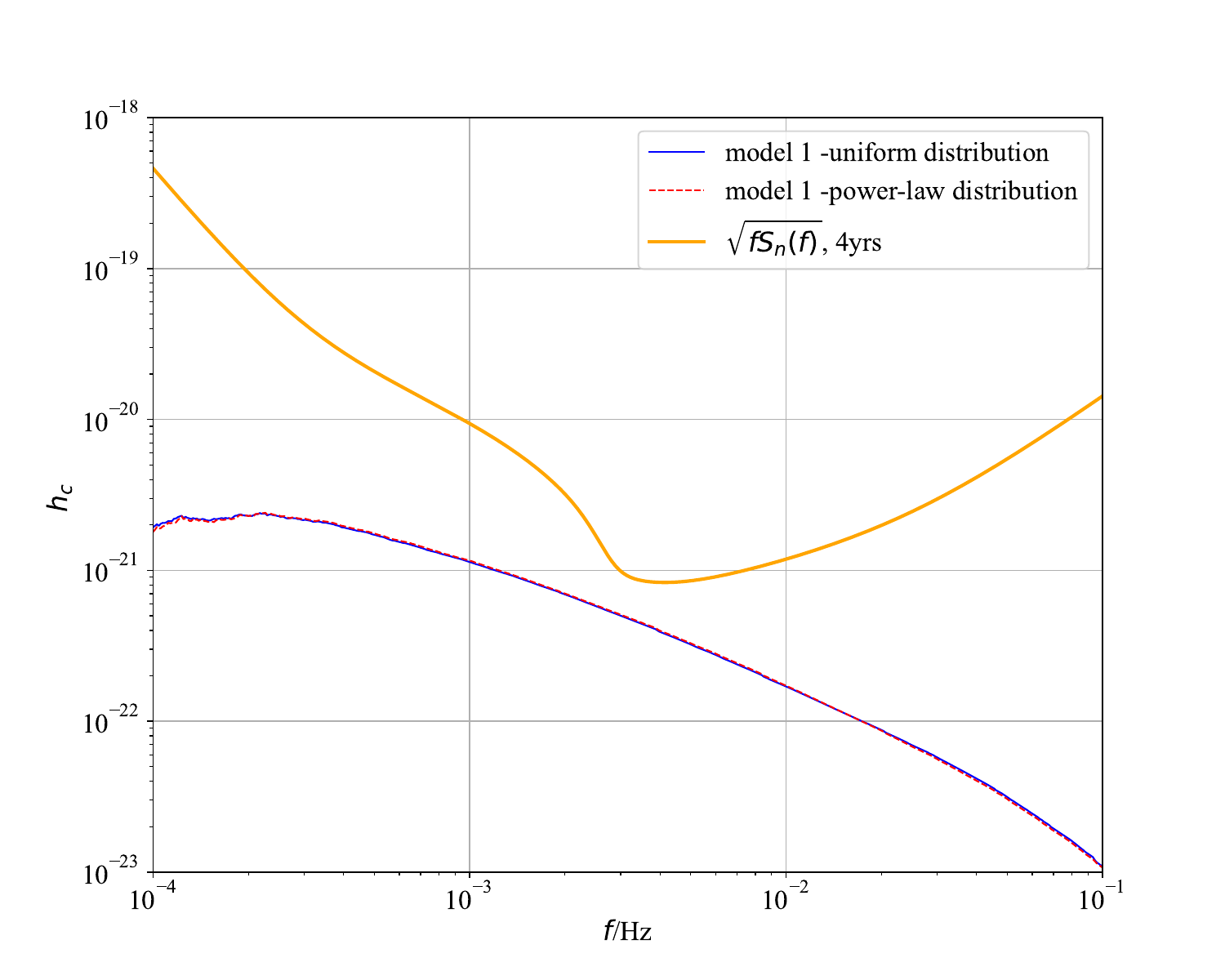}
    \caption{This figure shows the characteristic strain of the GWB corresponding to two different final eccentricity distributions. The blue solid line represents the case where the final eccentricity follows a uniform distribution, ranging from 0 to 0.2, while the red dashed line represents the case where the final eccentricity follows a power-law distribution with k = -0.9, also within the range of 0 to 0.2. It can be seen that the distribution of the final eccentricity does not significantly affect the characteristic strain of the GWB.}
    \label{fig:10}
\end{figure}

Furthermore, differences between assumed final eccentricity distributions—such as uniform vs. power-law—lead to distinctive features in the inferred $e_0$ distribution. When a uniform distribution in final eccentricity $e_{\text{final}} \in (0, 0.2)$ is assumed, systems are equally likely to be found near $e_{\text{final}} = 0$ and $e_{\text{final}} = 0.2$. Due to the nonlinear mapping between $e_0$ and $e_{\text{final}}$, systems with higher $e_{\text{final}}$ tend to originate from higher $e_0$, thereby producing more high-$e_0$ systems under the uniform assumption.
In contrast, if a power-law distribution is assumed for $e_{\text{final}}$, placing more weight on small eccentricities, then the backward evolution results in a larger number of systems with low $e_0$, while the contribution to the high-$e_0$ range is suppressed due to the relatively smaller number of systems with large $e_{\text{final}}$. As a result, compared to the uniform case, the power-law assumption yields more systems with $e_0 \in (0, 0.1)$ and fewer with $e_0 \in (0.8, 1)$. 
As shown in figure(\ref{fig9}), despite the large range in initial eccentricity, the resulting $h_c$ curves are nearly identical across the LISA-sensitive frequency band, with differences within a factor of a few at most. This confirms that the eccentricity of individual EMRIs—particularly their final values—has only a minor influence on the total GWB.
As a consequence, although the choice of final eccentricity distribution affects how we infer the initial dynamical conditions of EMRI systems, it does not significantly alter the predicted GWB, as shown in figure(\ref{fig:10}.)

\begin{figure}
\includegraphics[width=\columnwidth]{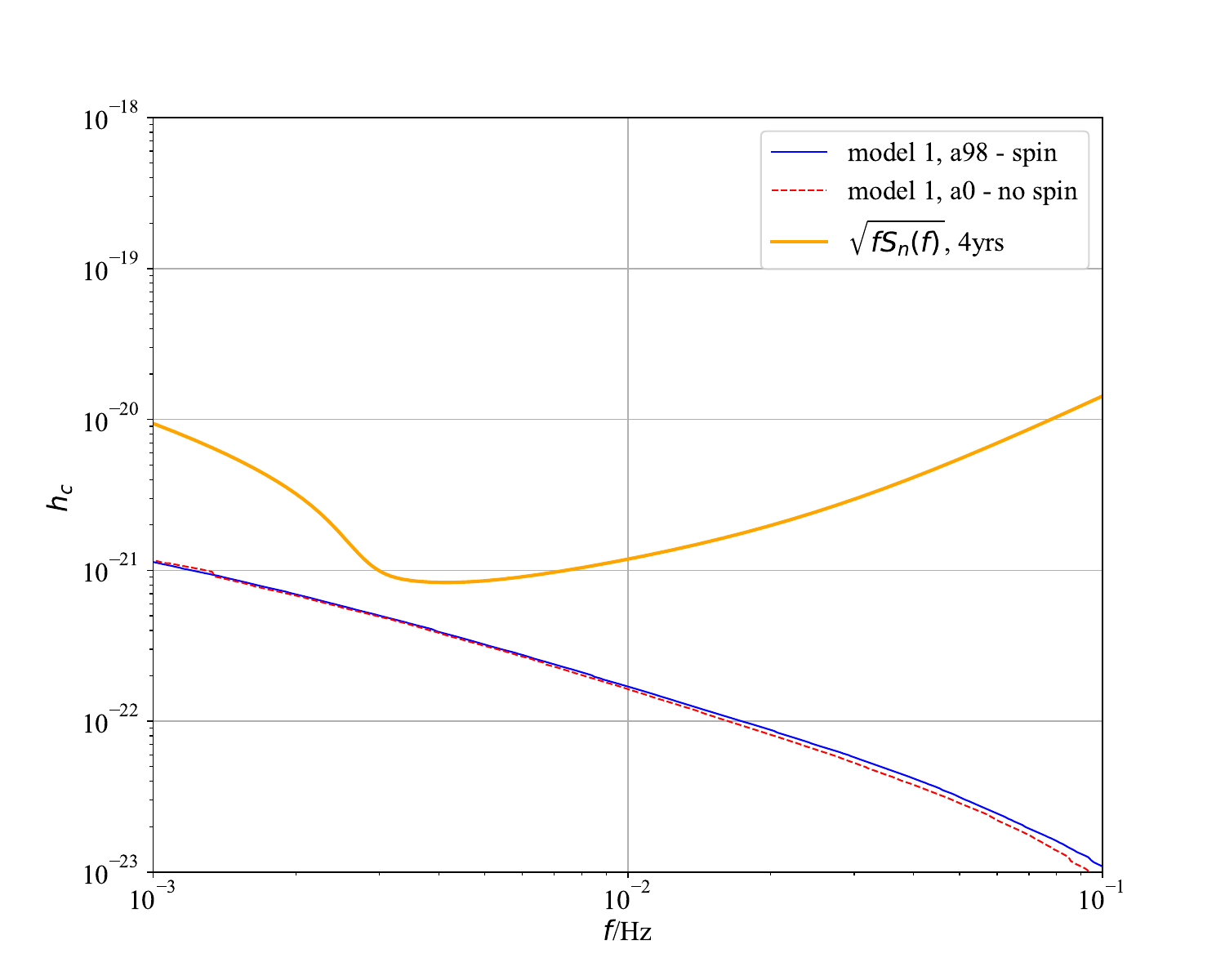}
    \caption{The blue solid line represents the case where the central black hole has spin, with a spin parameter  $a = 0.98$ , while the red dashed line represents the non-spinning case. It can be observed that in the low-frequency range, the effect of spin is not significant, but in the high-frequency range, a rapidly spinning black hole significantly enhances the GW radiation intensity. }
    \label{fig6}
\end{figure}

 The characteristic strain of the EMRI GWB increases if the spin of the central SMBH is higher shown in figure (\ref{fig6}). This enhancement arises from two key spin-related effects. First, the spin of a Kerr black hole leads to frame-dragging, which reduces the radius of the ISCO. As a result, compact objects can orbit closer to the SMBH, reaching regions of stronger gravitational potential and achieving higher orbital velocities. This leads to the emission of gravitational waves with greater intensity and higher frequencies.
Second, a rapidly spinning SMBH can also enhance the EMRI formation rate. For instance, the smaller ISCO and modified orbital dynamics increase the likelihood of compact objects being captured and retained in stable inspiral orbits, resulting in a greater number of EMRI events over time.
Among these two effects, the primary contributor to the GWB enhancement in our study is the increased EMRI formation rate, which is implemented in our population synthesis through a spin-dependent EMRI rate.
We note that our waveform modeling, based on Peters’ formula assuming non-spinning black holes, does not account for spin-induced modifications to individual EMRI waveforms. Developing a self-consistent framework that incorporates these spin effects in the future would enable a more accurate calculation of the EMRI gravitational wave background.

\begin{figure}
\includegraphics[width=\columnwidth]{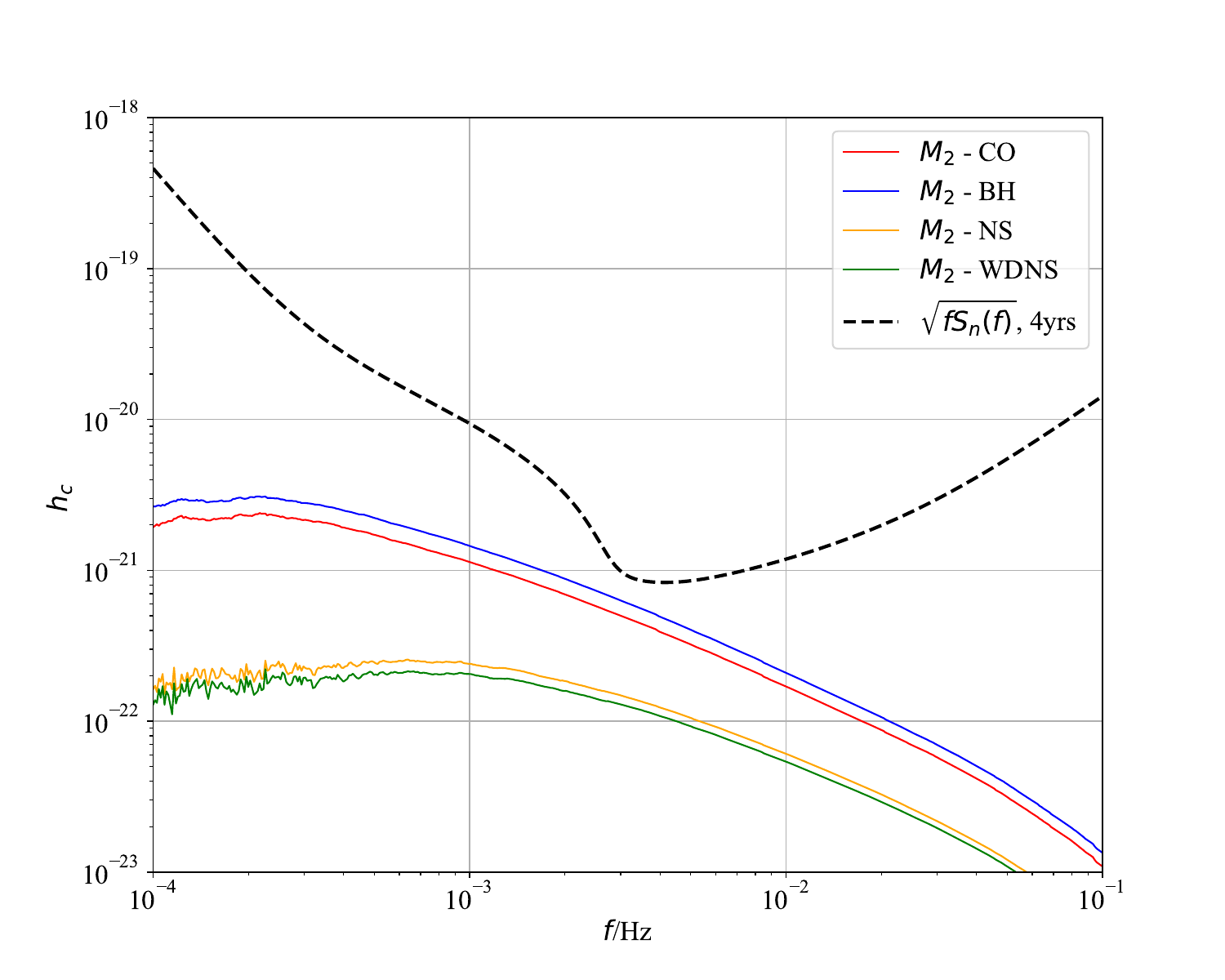}
    \caption{This figure illustrates the impact of different compact objects on the gravitational wave signal. The term “CO” refers to all compact objects, which include black holes (BH), neutron stars (NS), and white dwarf-neutron star (WD-NS) systems, with masses in the range of 1–50  $M_{\odot}$ . When the compact object is a BH , whose mass range is 5–50  $M_{\odot}$, the gravitational wave signal is the strongest, especially in the low-frequency range. In contrast, signals from NS systems, with masses n the range of 1-2  $M_{\odot}$ , and WD-NS systems, with masses ranging from 0.4 to 2  $M_{\odot}$ , are weaker. LISA’s optimal frequency range for detecting these signals is around  $10^{-2}$  Hz, making gravitational wave signals from systems with black hole companions more easily detectable.
 }
    \label{7}
\end{figure}

We also test the influence of varying CO masses on the GW background. As shown in fig(\ref{7}), when the CO is a black hole, the characteristic strain is the highest, especially in the low-frequency range. But NS and WD-NS systems produce weaker signals. LISA’s detection sensitivity occurs around $10^{-2}$ Hz, making gravitational wave signals from systems with black hole more detectable.

The lower gravitational wave radiation intensity from NS and WD-NS systems at low frequencies is due to the nature of EMRI systems, where the GW radiation primarily arises from the gradual inspiral of the CO as it spirals inward toward the central black hole. Because NSs have relatively low mass, their orbital shrinkage rate with the central black hole is slower, requiring more time to evolve into higher frequency ranges, which correspond to tighter orbits. At low frequencies, the orbits remain relatively loose, resulting in lower-frequency gravitational wave emissions, and the radiation intensity has not yet reached higher levels. In contrast, massive objects like black holes generate stronger gravitational wave radiation in this frequency range.

Finally, the choice of the $M$--$\sigma$ relation, which connects the mass of the central SMBH to the stellar velocity dispersion of its host galaxy, can influence the predicted EMRI event rate and consequently the GWB amplitude. In this work, we adopt the widely used relation from G\"ultekin et al.\ (2009)\cite{Gultekin2009} as a baseline model. However, alternative relations—such as those proposed by Kormendy \& Ho (2013)\cite{Kormendy2013} or Graham \& Scott (2013)\cite{Graham2013}—predict systematically different SMBH masses for the same host velocity dispersion, which could shift the SMBH mass function and affect EMRI rates.

These differences impact both the abundance of SMBHs in certain mass ranges and the timescales for stellar cusp regeneration. For instance, the cusp regeneration timescale $t_{\mathrm{cusp}}$ is typically shorter under the Graham \& Scott (2013) relation (e.g., $\sim 2$ Gyr), and longer under Kormendy \& Ho (2013) (e.g., $\sim 10$ Gyr), compared to $\sim 6$ Gyr for G\"ultekin et al.\ (2009). A shorter $t_{\mathrm{cusp}}$ implies more efficient repopulation of the central regions with compact objects, potentially enhancing the EMRI formation rate and the resulting GWB.

Although variations in the $M$--$\sigma$ relation can affect the overall amplitude of the GWB—typically within a factor of a few—they do not significantly alter its spectral shape. Since this source of uncertainty originates from current empirical scatter among observational studies, future refinement of galactic scaling relations will be important for improving GWB predictions. A more systematic investigation of this dependence will be carried out in future work.

 Our study incorporates a more detailed population model, although our methodology is similar to the approaches used in Bonetti \& Sesana (2020)\cite{Bonetti2020} to calculate the EMRI GWB.
In particular,  (1) we  distinguish different types of compact object companions such as stellar-mass BHs (5--50\,$M_\odot$), NSs (1--2\,$M_\odot$), and WD-NS systems (0.4--2\,$M_\odot$); (2) we thus compute the contribution to the GWB of each of them; (3) we also examine various galactic environments and a broader range of SMBH spins. 
In addition, Bonetti \& Sesana (2020) assumed a uniform distribution for final eccentricity in the range of$e_{\mathrm{final}} \in (0, 0.2)$. In our analysis, we explore both uniform and power-law distributions to test the robustness of the GWB results. Our extensions allow us to capture a wider variety of astrophysical scenarios and yield more flexible and detailed predictions for the EMRI GWB.

\section{Conclusion}
We have conducted an extensive analysis of the stochastic GWB generated by a cosmological population of EMRIs within the LISA frequency band. Through our simulations and calculations of the GWB from EMRIs, we have reached the following conclusions :
(1) The GWB from EMRIs, especially those simulated using Barausse12, has a high probability of being detected in the $10^{-3} - 10^{-2} ${Hz} frequency band, which lies above LISA’s sensitivity curve . (2) Our findings indicate that variations in eccentricity at the end stage of the inspiral process have little significant effect on the overall strength of the GWB, so the influence of the final eccentricity distribution can be neglected in the later calculations. (3) The rapidly spinning SMBH slightly increases the GW characteristic strain compared to a non-spinning Schwarzschild black hole, with spin increasing the strain by approximately 0.1 orders of magnitude, particularly in the $10^{-2} - 10^{-1} \text{Hz}$ frequency band. (4) The characteristic strain of the GWB from EMRIs is strongly influenced by the masses of COs, with BHs as COs generating a GW signal intensity roughly one order of magnitude higher than that produced by NSs or WDs, especially in the low-frequency range of $10^{-4} - 10^{-3}\text{Hz}$.

In this work, we go beyond single EMRI events, adopting instead a broader population-based perspective on the EMRI GWB. By investigating the statistical properties of EMRI systems under various astrophysical parameters, we extract information about black holes and their galactic environments. However, we did not fully include the detailed dynamics of single EMRI systems, such as how a black hole’s spin or the dark matter around it might affect each orbit\cite{Li2022}. Single EMRI events tend to emphasize their dynamical features, whereas the EMRI GWB emphasizes their statistical characteristics. Addressing these issues will provide a more precise theoretical framework for GWB detection, offering essential insights for future GW experiments. 
Besides, there are some new theories regarding the gravitational wave background from EMRIs, such as a novel EMRI production channel known as the “cliffhanger” EMRI, which can dominate volumetric EMRI rates and potentially increase EMRI rates by an
order of magnitude\cite{Qunbar2023}. Our future work can be developed based on these new theories and will also lead to a deeper understanding of the GWB from EMRIs.

\section*{Acknowledgements}
We would like to thank the referees for their valuable comments and inputs which significantly improved the original manuscript. 
This work is supported by National Key R\&D Program of China (2020YFC2201400).  


\end{document}